\documentclass[twocolumn,amsmath,amssymb,pra]{revtex4}
\usepackage[dvips]{graphicx}
\usepackage{dcolumn}
\usepackage{bm}
\newcounter{one}
\begin{document}

\title{Experimental generation of four-mode continuous-variable cluster states}

\author
{Mitsuyoshi Yukawa$^{1,2}$, Ryuji Ukai$^{1,2}$, Peter van Loock$^{3}$, and Akira Furusawa$^{1,2}$}

\affiliation{$^{1}$Department of Applied Physics, School of Engineering, The University of Tokyo,\\ 7-3-1 Hongo, Bunkyo-ku, Tokyo 113-8656, Japan\\
$^{2}$CREST, Japan Science and Technology (JST) Agency, 5, Sanbancho, Chiyoda-ku, Tokyo 102-0075, Japan\\
$^{3}$Optical Quantum Information Theory Group, Institute of Theoretical Physics \Roman{one} and Max-Planck Research Group, Institute of Optics, Information and Photonics, Universit$\ddot{a}$t Erlangen-N$\ddot{u}$rnberg, Staudtstr. 7/B2, 91058 Erlangen, Germany}

\begin{abstract}
Continuous-variable Gaussian cluster states
are a potential resource for universal quantum computation.
They can be efficiently and unconditionally
built from sources of squeezed light using beam splitters.
Here we report on the generation of three different
kinds of continuous-variable four-mode cluster states.
In our realization, the resulting cluster-type correlations
are such that no corrections other than simple phase-space displacements
would be needed when quantum information propagates through
these states. At the same time, the inevitable imperfections
from the finitely squeezed resource states and from additional
thermal noise are minimized, as no antisqueezing components
are left in the cluster states.
\end{abstract}

\maketitle

\section{Introduction}

The one-way quantum computation model \cite{Raussendorf01} is a conceptually
interesting alternative to the standard circuit model.
One-way quantum computation can be carried out with the help of
a specially prepared entangled state, a so-called cluster state.
Quantum information is then processed solely through measurements on the cluster
state. Universal quantum gates can be achieved by choosing different measurement bases
and using feedforward.

Small-scale discrete-variable (DV) cluster computations have been demonstrated
already with single photons \cite{Walther05}. An alternative approach to, in particular,
an optical implementation of cluster-based quantum computation is based upon
continuous variables. Continuous-variable (CV) one-way quantum computation
\cite{Menicucci06} relies upon CV cluster states \cite{Zhang06}.
Various ways for generating CV cluster-type states are known
\cite{Menicucci06,Zhang06,Menicucci07,Zaidi08,Peter07}.
The method employed in this work is to combine beams of off-line squeezed light at beam splitters \cite{Peter07}.
Once CV cluster states are available, quantum information encoded into an optical mode
can be manipulated via homodyne and non-Gaussian measurements on the CV cluster
\cite{Menicucci06,Peter07-2}. As the elementary routine for cluster computation is
quantum teleportation, first steps towards the implementation of CV cluster computation
have been achieved already,
including the realization of a quantum teleportation network \cite{Yonezawa04},
entanglement swapping \cite{Takei05}, teleportation of squeezing \cite{Yonezawa07},
sequential quantum teleportation \cite{Yonezawa07a}, and
high-fidelity teleportation \cite{Yukawa08}.
Here we report on the optical creation of four-mode CV cluster states suitable for
small-scale implementations of CV one-way quantum computation.

The quadrature correlations of cluster-type states are such that in the limit of
infinite squeezing, the states become zero eigenstates of a set of quadrature combinations,
\vspace{-10mm}
\begin{center}
\begin{equation}
\biggl(\hat{p}_{a}-\sum\limits_{b\in N_{a}}\hat{x}_{b}\biggr)\to 0\;,\;\;\;\forall a\in G\;.
\label{ClusterDefinition}
\end{equation}
\end{center}
Here, $\hat{x}_{a}$ and $\hat{p}_{a}$ correspond to ``position'' and ``momentum'' operators for an
optical mode $a$ with annihilation operator $\hat{a}_{a}=\hat{x}_{a}+i\hat{p}_{a} (\hbar=\frac{1}{2})$.
Every mode $a\in G$ represents a vertex of the graph $G$ (representing the cluster or graph state).
The modes $b\in N_{a}$ are the nearest neighbors of mode $a$.

A possible way to obtain CV cluster states is to entangle a corresponding number of optical
modes, each initially in a squeezed state, through quantum nondemolition (QND) interactions
\cite{Zhang06}, in analogy to the creation of qubit cluster states via controlled sign gates.
We may refer to this specific type of cluster states as canonical cluster states \cite{Peter07}.
Experimentally, the optical CV QND gates for every single link of the cluster state
can be realized with two beam splitters and two on-line squeezers
\cite{Braunstein05} for each link. Alternatively, the initial squeezing transformations can be absorbed
into the entire QND network; after Bloch-Messiah reduction \cite{Braunstein05}
only off-line squeezed states and linear optics are effectively
needed then to produce a canonical cluster state \cite{Peter07}.

In another approach for building CV cluster-type states from squeezed light using linear optics
\cite{Peter07}, the beam splitter network is carefully chosen such that, by construction,
all antisqueezing components are completely eliminated in the output operator combinations,
$\hat{p}_{a}-\Sigma_{b\in N_{a}}\hat{x}_{b}$; hence these combinations, being proportional
to the squeezing factor, $\hat{p}_{a}-\Sigma_{b\in N_{a}}\hat{x}_{b}\propto e^{-r}$,
automatically satisfy the conditions of Eq.~(\ref{ClusterDefinition}) in the limit of infinite squeezing
$r\to\infty$. Moreover, generating cluster-type states in this way requires smaller degrees
of input squeezing than needed for making the canonical states with the same quality of correlations
\cite{Peter07}.

The complete removal of antisqueezing components is particularly beneficial,
as in the actual experiment, the antisqueezing levels are typically greater than the squeezing levels
due to experimental imperfections such as losses and fluctuations in the phase locking.
By employing the above-mentioned method for eliminating the antisqueezing components
in our experiment,
we can observe that the single-mode squeezing levels
of the input states before the generation of the cluster states are effectively reproduced in the multi-mode
squeezing levels of the resulting cluster states. This is in contrast to the experiments of
Refs.~\cite{Su07,Tan08}, where the antisqueezing components are not completely suppressed.
Another advantage of our approach here is that the resulting quadrature correlations
are precisely those occurring in the excess noise terms when quantum information propagates
through a CV Gaussian cluster state \cite{Menicucci06}. Suppressing this excess noise efficiently
means reducing the errors in cluster-based quantum computations.

\section{Gaussian Four-Mode Cluster States} \label{SecGeneFourModeCluster}

\begin{figure}[htbp]
  \begin{center}
    \includegraphics[width=0.9\linewidth]{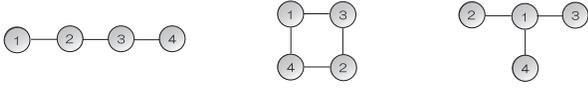}
  \end{center}
  \vspace{-4mm}
  \caption{The created four-mode cluster states.
  Each cluster node, corresponding to an optical mode, is represented by a circle.
  Neighboring nodes are connected by lines.}
  \label{fig:cluster_states.eps}
\end{figure}

According to the efficient method proposed in Ref.~\cite{Peter07}, we created
three kinds of four-mode CV cluster states in this experiment, including a linear cluster state,
a square cluster state, and a T-shape cluster state (see Fig.\ref{fig:cluster_states.eps}).

In general, four-mode CV graph states can be built from four off-line squeezed states using
up to six beam splitters \cite{Peter07,Reck94}. For the linear four-mode cluster state,
three beam splitters have been shown to be sufficient \cite{Peter07}. Similarly, only
three beam splitters are needed to produce the square and the T-shape cluster states.
The optical setups are described in detail below.

In order to create these cluster states,
four $p$-squeezed states, with mode operators
$\hat{a}_{i}=e^{+r_{i}}\hat{x}_{i}^{(0)}+ie^{-r_{i}}\hat{p}_{i}^{(0)}(i=1\hdots 4)$, are prepared first.
A superscript (0) denotes initial vacuum modes and $r_{i}$ is the squeezing parameter of the $i$th mode.
For a unitary matrix $U$ representing a sequence of beam splitters, the output mode operators
$\hat{a}_{i}'$ can be obtained according to $\hat{a}_{i}'=\Sigma_{j}U_{ij}\hat{a}_{j}$.

A possible solution for the matrix $U_{\mathrm{L}}$, giving the linear cluster state, is \cite{Peter07}
\vspace{-5mm}\begin{center}
\begin{equation}
\begin{split}
U_{\mathrm{L}}=&
\left(
\begin{array}{cccc}
\frac{1}{\sqrt{2}} & \frac{1}{\sqrt{10}} & \frac{2i}{\sqrt{10}} & 0\\
\frac{i}{\sqrt{2}} & -\frac{i}{\sqrt{10}} & \frac{2}{\sqrt{10}} & 0\\
0 & -\frac{2}{\sqrt{10}} & \frac{i}{\sqrt{10}} & \frac{i}{\sqrt{2}}\\
0 & -\frac{2i}{\sqrt{10}} & -\frac{1}{\sqrt{10}} & \frac{1}{\sqrt{2}}
\end{array}
\right).
\end{split}
\end{equation}
\end{center}

With this matrix, the quadrature quantum correlations of the output state become
\vspace{-5mm}
\begin{center}
\begin{equation}
\begin{split}
\hat{p}_{\mathrm{L}1}-&\hat{x}_{\mathrm{L}2}=\sqrt{2}e^{-r_{1}}\hat{p}_{1}^{(0)}\\
\hat{p}_{\mathrm{L}2}-\hat{x}_{\mathrm{L}1}-\hat{x}_{\mathrm{L}3}=&\sqrt{\frac{5}{2}}e^{-r_{3}}\hat{p}_{3}^{(0)}+\frac{1}{\sqrt{2}}e^{-r_{4}}\hat{p}_{4}^{(0)}\\
\hat{p}_{\mathrm{L}3}-\hat{x}_{\mathrm{L}2}-\hat{x}_{\mathrm{L}4}=&\frac{1}{\sqrt{2}}e^{-r_{1}}\hat{p}_{1}^{(0)}-\sqrt{\frac{5}{2}}e^{-r_{2}}\hat{p}_{2}^{(0)}\\
\hat{p}_{\mathrm{L}4}-&\hat{x}_{\mathrm{L}3}=\sqrt{2}e^{-r_{4}}\hat{p}_{4}^{(0)}.
\end{split}
\end{equation}
\end{center}
All these linear combinations are proportional to the squeezing factors,
approaching zero in the limit of infinite squeezing.
Hence, the output state is a linear four-mode cluster state,
in agreement with Eq.~(\ref{ClusterDefinition}).

The matrix $U_{\mathrm{L}}$ can be decomposed into $U_{\mathrm{L}}=F_{4}S_{12}F_{1}^{\dagger}B_{34}^{+}(1/\sqrt{2})B_{21}^{+}(1/\sqrt{2})B_{23}^{-}(1/\sqrt{5})F_{3}F_{4}$.
Here, $F_{k}$ denotes the Fourier transform ($90^\circ$ rotation in phase space) of mode $k$, $\hat{a}_{k}\to i\hat{a}_{k}$.
$B_{ij}^{\pm}(t)$ corresponds to a beam splitter
transformation of modes $i$ and $j$ with transmittance parameter $t$;
it is equivalent to the four-mode identity matrix
except for $(B_{ij}^{\pm})_{ii}=t$, $(B_{ij}^{\pm})_{ij}=\sqrt{1-t^{2}}$,
$(B_{ij}^{\pm})_{ji}=\pm\sqrt{1-t^{2}}$ and $(B_{ij}^{\pm})_{jj}=\mp t$.
$S_{ij}$ is the swapping operation of modes $i$ and $j$.
As a result, two symmetric beam splitters and one 1:4 beam splitter can be used.

Let us consider next the square cluster state. A possible solution
for the unitary matrix $U_{\mathrm{S}}$ is
\vspace{-5mm}\begin{center}
\begin{equation}
\begin{split}
U_{\mathrm{\mathrm{S}}}=&
\left(
\begin{array}{cccc}
-\frac{1}{\sqrt{2}} & -\frac{1}{\sqrt{10}} & -\frac{2i}{\sqrt{10}} & 0\\
\frac{1}{\sqrt{2}} & -\frac{1}{\sqrt{10}} & -\frac{2i}{\sqrt{10}} & 0\\
0 & -\frac{2i}{\sqrt{10}} & -\frac{1}{\sqrt{10}} & -\frac{1}{\sqrt{2}}\\
0 & -\frac{2i}{\sqrt{10}} & -\frac{1}{\sqrt{10}} & \frac{1}{\sqrt{2}}
\end{array}
\right).
\end{split}
\end{equation}
\end{center}

The correlations of the output state are
\vspace{-5mm}
\begin{center}
\begin{equation}
\begin{split}
\hat{p}_{\mathrm{S} 1}-\hat{x}_{\mathrm{S} 3}-\hat{x}_{\mathrm{S} 4}&=-\frac{1}{\sqrt{2}}e^{-r_{1}}\hat{p}_{1}^{(0)}-\sqrt{\frac{5}{2}}e^{-r_{2}}\hat{p}_{2}^{(0)}\\
\hat{p}_{\mathrm{S} 2}-\hat{x}_{\mathrm{S} 3}-\hat{x}_{\mathrm{S} 4}&=\frac{1}{\sqrt{2}}e^{-r_{1}}\hat{p}_{1}^{(0)}-\sqrt{\frac{5}{2}}e^{-r_{2}}\hat{p}_{2}^{(0)}\\
\hat{p}_{\mathrm{S} 3}-\hat{x}_{\mathrm{S} 1}-\hat{x}_{\mathrm{S} 2}&=-\sqrt{\frac{5}{2}}e^{-r_{3}}\hat{p}_{3}^{(0)}-\frac{1}{\sqrt{2}}e^{-r_{4}}\hat{p}_{4}^{(0)}\\
\hat{p}_{\mathrm{S} 4}-\hat{x}_{\mathrm{S} 1}-\hat{x}_{\mathrm{S} 2}&=-\sqrt{\frac{5}{2}}e^{-r_{3}}\hat{p}_{3}^{(0)}+\frac{1}{\sqrt{2}}e^{-r_{4}}\hat{p}_{4}^{(0)},
\end{split}
\end{equation}
\end{center}
all of which approach zero in the limit of infinite squeezing.

We find that the matrix $U_{\mathrm{S}}$
is equal to $U_{\mathrm{add}}U_{\mathrm{L}}$, with $U_{\mathrm{add}}=diag(-1,-i,i,1)$.
Thus, the square cluster state can be obtained from the linear cluster state via
local Fourier transforms.
In the experiment, the local Fourier transforms can be easily achieved
by changing the locking phase of the local oscillator beams.
Therefore, the optical setup for the square cluster state is more or less
identical to that for the linear cluster state. Using the identities,
$\hat{p}_{\mathrm{S} 1}-\hat{x}_{\mathrm{S} 3}-\hat{x}_{\mathrm{S} 4}=-\hat{p}_{\mathrm{L}1}+\hat{p}_{\mathrm{L}3}-\hat{x}_{\mathrm{L}4},
\hat{p}_{\mathrm{S} 2}-\hat{x}_{\mathrm{S} 3}-\hat{x}_{\mathrm{S} 4}=-\hat{x}_{\mathrm{L}2}+\hat{p}_{\mathrm{L}3}-\hat{x}_{\mathrm{L}4},
\hat{p}_{\mathrm{S} 3}-\hat{x}_{\mathrm{S} 1}-\hat{x}_{\mathrm{S} 2}=\hat{x}_{\mathrm{L}1}-\hat{p}_{\mathrm{L}2}+\hat{x}_{\mathrm{L}3},
\hat{p}_{\mathrm{S} 4}-\hat{x}_{\mathrm{S} 1}-\hat{x}_{\mathrm{S} 2}=\hat{x}_{\mathrm{L}1}-\hat{p}_{\mathrm{L}2}+\hat{p}_{\mathrm{L}4}$,
the measurement of the correlations of the linear cluster state and the square cluster state can be performed in a single experiment.

Finally, the T-shape cluster state can be obtained from four $p$-squeezed states followed by a unitary transform,
\vspace{-5mm}
\begin{center}
\begin{equation}
U_{\mathrm{T}}=\left(
  \begin{array}{cccc}
\frac{i}{\sqrt{2}} & \frac{1}{2} & \frac{i}{2} & 0\\
\frac{1}{\sqrt{2}} & \frac{i}{2} & -\frac{1}{2} & 0\\
0 & \frac{i}{2} & \frac{1}{2} & \frac{1}{\sqrt{2}}\\
0 & \frac{i}{2} & \frac{1}{2} & -\frac{1}{\sqrt{2}}
  \end{array}
 \right).
\end{equation}
\end{center}

This can be also decomposed into $U_{\mathrm{T}}=F_{1}^{\dagger}B_{34}^{+}(1/\sqrt{2})B_{21}^{+}(1/\sqrt{2})B_{32}^{-}(1/\sqrt{2})F_{2}$.
Thus, this time, the optical setup has to be modified, but three beam splitters are still sufficient.
The quantum correlations of the output state are
\vspace{-7mm}
\begin{center}
\begin{equation}
\begin{split}
\hat{p}_{\mathrm{T}1}&-\hat{x}_{\mathrm{T}2}-\hat{x}_{\mathrm{T}3}-\hat{x}_{\mathrm{T}4}=2e^{-r_{2}}\hat{p}_{2}^{(0)}\\
&\hat{p}_{\mathrm{T}2}-\hat{x}_{\mathrm{T}1}=\sqrt{2}e^{-r_{1}}\hat{p}_{1}^{(0)}\\
\hat{p}_{\mathrm{T}3}-\hat{x}_{\mathrm{T}1}=\frac{1}{\sqrt{2}}&e^{-r_{2}}\hat{p}_{2}^{(0)}+e^{-r_{3}}\hat{p}_{3}^{(0)}+\frac{1}{\sqrt{2}}e^{-r_{4}}\hat{p}_{4}^{(0)}\\
\hat{p}_{\mathrm{T}4}-\hat{x}_{\mathrm{T}1}=\frac{1}{\sqrt{2}}&e^{-r_{2}}\hat{p}_{2}^{(0)}+e^{-r_{3}}\hat{p}_{3}^{(0)}-\frac{1}{\sqrt{2}}e^{-r_{4}}\hat{p}_{4}^{(0)}.
\end{split}
\end{equation}
\end{center}
They all approach zero in the limit of infinite squeezing.
Hence, the output state is a T-shape cluster state, according to Eq.~(\ref{ClusterDefinition}).
Note that up to local Fourier transforms, the T-shape cluster state is equivalent
to a four-mode GHZ-type state \cite{Peter00}.

\begin{figure}[b]
  \begin{center}
    \includegraphics[width=0.75\linewidth]{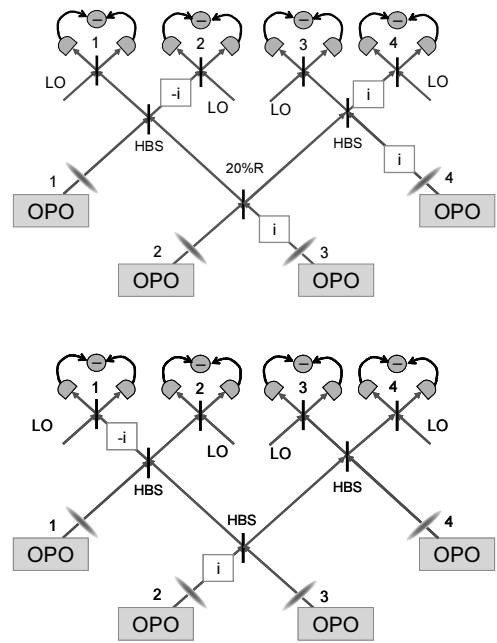}
  \end{center}
  \caption{The schematic of the optical setups to create the linear cluster state (top) and T-shape cluster state (bottom). Four squeezed states are generated by OPOs (optical parametric oscillators). HBS is half beam splitter and 20$\%$ is 1:4 beam splitter. Boxes including $i$ are Fourier transforms ($90^{\circ}$ rotations in phase space), and $-i$ is a $-90^{\circ}$ rotation. LO is local oscillator for homodyne detection of the output states.}
    \label{fig:setup.eps}
\end{figure}

\section{Experimental Implementation}

The schematic of our optical setups is shown in Fig.\ref{fig:setup.eps}.
We use a continuous-wave Ti:sapphire laser (Coherent MBR110, $\lambda$=860nm) as a light source.
In order to generate squeezed states, optical parametric oscillators (OPOs) are used via
optical degenerate parametric downconversion.
Periodically poled $\mathrm{KTiOPO_{4}}$ (PPKTP) crystals are employed as nonlinear optical media.
Each OPO is pumped by a second harmonic beam obtained from a cavity
which contains a potassium niobate ($\mathrm{KNbO_{3}}$) crystal for second harmonic generation.
The pump powers range from 76mW to 96mW.

Weak coherent beams are also injected into the OPOs and the emitted beams are set to 2$\mu$W.
On each beam, phase modulations are applied for locking, with 140kHz for OPO2, 210kHz for OPO3,
and 98kHz for OPO1 and OPO4.
In this experiment, a 1MHz sideband is chosen for the measurements,
so the phase modulations do not affect these measurements.

An output state is measured via homodyne detection with a strong beam around 5mW used as a local oscillator.
The homodyne detector gives a voltage signal of the measurement result.
After electronically combining the outputs of the homodyne detectors,
the signals of the correlations can be obtained.
The signals are sent to a spectrum analyzer in order to get the mesurement data.

\begin{figure}[htb]
  \begin{center}
    \includegraphics[width=\linewidth]{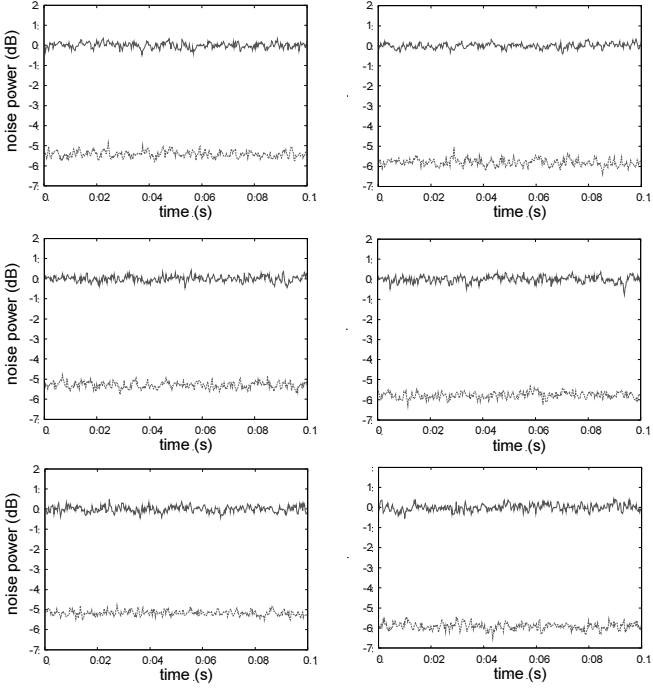}
  \end{center}
  \caption{In all graphs, the measurement variances without squeezing (upper) and with squeezing (lower) are shown. The graphs of the first row are the results of $\langle[\Delta (\hat{p}_{\mathrm{L}1}-\hat{x}_{\mathrm{L}2})]^{2}\rangle$ and $\langle[\Delta (\hat{p}_{\mathrm{L}2}-\hat{x}_{\mathrm{L}1}-\hat{x}_{\mathrm{L}3})]^{2}\rangle=\langle[\Delta (\hat{p}_{\mathrm{S} 3}-\hat{x}_{\mathrm{S} 1}-\hat{x}_{\mathrm{S} 2})]^{2}\rangle$. The ones of the second row are for $\langle[\Delta (\hat{p}_{\mathrm{L}3}-\hat{x}_{\mathrm{L}2}-\hat{x}_{\mathrm{L}4})]^{2}\rangle=\langle[\Delta (\hat{p}_{\mathrm{S} 2}-\hat{x}_{\mathrm{S} 3}-\hat{x}_{\mathrm{S} 4})]^{2}\rangle$ and $\langle[\Delta (\hat{p}_{\mathrm{L}4}-\hat{x}_{\mathrm{L}3})]^{2}\rangle$. The ones of the third row are for $\langle[\Delta (\hat{p}_{\mathrm{S} 1}-\hat{x}_{\mathrm{S} 3}-\hat{x}_{\mathrm{S} 4})]^{2}\rangle$ and $\langle[\Delta (\hat{p}_{\mathrm{S} 4}-\hat{x}_{\mathrm{S} 1}-\hat{x}_{\mathrm{S} 2})]^{2}\rangle$. The measurement frequency is 1MHz, resolution bandwidth is 30kHz, and video bandwidth is 300Hz. All results are obtained with 20 times averaging.}
  \label{fig:LinearResult.eps}
\end{figure}

\begin{figure}[htb]
  \begin{center}
    \includegraphics[width=\linewidth]{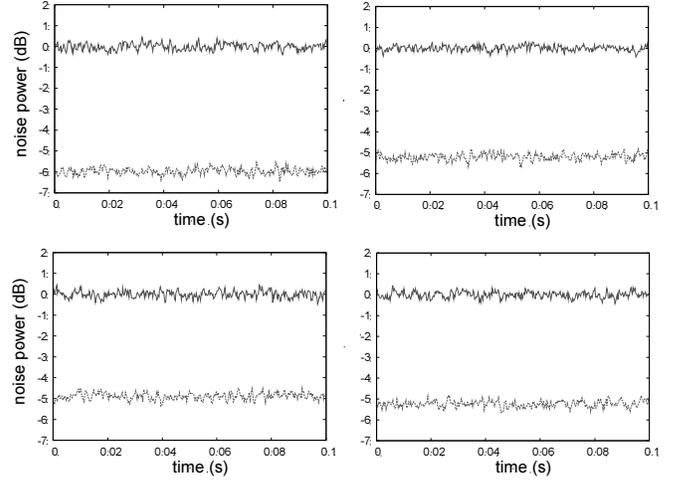}
  \end{center}
  \caption{The upper ones are $\langle[\Delta (\hat{p}_{\mathrm{T}1}-\hat{x}_{\mathrm{T}2}-\hat{x}_{\mathrm{T}3}-\hat{x}_{\mathrm{T}4})]^{2}\rangle$ and $\langle[\Delta (\hat{p}_{\mathrm{T}2}-\hat{x}_{\mathrm{T}1})]^{2}\rangle$. The bottom ones are $\langle[\Delta (\hat{p}_{\mathrm{T}3}-\hat{x}_{\mathrm{T}1})]^{2}\rangle$ and $\langle[\Delta (\hat{p}_{\mathrm{T}4}-\hat{x}_{\mathrm{T}1})]^{2}\rangle$. The conditions of the measurements are the same as in Fig.~\ref{fig:LinearResult.eps}}
    \label{fig:T-shapeResult.eps}
\end{figure}

Figs.~\ref{fig:LinearResult.eps} and \ref{fig:T-shapeResult.eps} show the results of the measurements.
Theoretically, the correlations should be proportional to the squeezing levels.
Every graph shows the results with squeezing and without squeezing.
For the linear cluster state and the square cluster state, $\langle[\Delta (\hat{p}_{\mathrm{L}1}-\hat{x}_{\mathrm{L}2})]^{2}\rangle=-5.4\pm0.2\mathrm{dB}$, $\langle[\Delta (\hat{p}_{\mathrm{L}2}-\hat{x}_{\mathrm{L}1}-\hat{x}_{\mathrm{L}3})]^{2}\rangle=\langle[\Delta (\hat{p}_{\mathrm{S} 3}-\hat{x}_{\mathrm{S} 1}-\hat{x}_{\mathrm{S} 2})]^{2}\rangle=-5.8\pm0.2\mathrm{dB}$, $\langle[\Delta (\hat{p}_{\mathrm{L}3}-\hat{x}_{\mathrm{L}2}-\hat{x}_{\mathrm{L}4})]^{2}\rangle=\langle[\Delta (\hat{p}_{\mathrm{S} 2}-\hat{x}_{\mathrm{S} 3}-\hat{x}_{\mathrm{S} 4})]^{2}\rangle=-5.3\pm0.2\mathrm{dB}$, $\langle[\Delta (\hat{p}_{\mathrm{L}4}-\hat{x}_{\mathrm{L}3})]^{2}\rangle=-5.8\pm0.2\mathrm{dB}$, $\langle[\Delta (\hat{p}_{\mathrm{S} 1}-\hat{x}_{\mathrm{S} 3}-\hat{x}_{\mathrm{S} 4})]^{2}\rangle=-5.2\pm0.2\mathrm{dB}$, $\langle[\Delta (\hat{p}_{\mathrm{S} 4}-\hat{x}_{\mathrm{S} 1}-\hat{x}_{\mathrm{S} 2})]^{2}\rangle=-5.9\pm0.2\mathrm{dB}$ are obtained. We point out again that only six measurements are sufficient
to detect the eight correlations of these two states.

For the T-shape cluster state, the results of the measurements are $\langle[\Delta (\hat{p}_{\mathrm{T}1}-\hat{x}_{\mathrm{T}2}-\hat{x}_{\mathrm{T}3}-\hat{x}_{\mathrm{T}4})]^{2}\rangle=-6.0\pm0.2\mathrm{dB}$, $\langle[\Delta (\hat{p}_{\mathrm{T}2}-\hat{x}_{\mathrm{T}1})]^{2}\rangle=-5.2\pm0.2\mathrm{dB}$, $\langle[\Delta (\hat{p}_{\mathrm{T}3}-\hat{x}_{\mathrm{T}1})]^{2}\rangle=-4.9\pm0.2\mathrm{dB}$ and $\langle[\Delta (\hat{p}_{\mathrm{T}4}-\hat{x}_{\mathrm{T}1})]^{2}\rangle=-5.2\pm0.2\mathrm{dB}$.


Various ways for constructing multi-party entanglement witnesses (i.e., observables for detecting the presence of
multi-party entanglement) are known in the regime of CV multi-mode states \cite{Inseparability, Hyllus06}.
We employ the method of Ref.~\cite{Inseparability} using a set of sufficient conditions for the full inseparability
of a multi-mode state which can be easily tested with our experimental results.
The corresponding inequalities to be satisfied are shown below. Recall that
if the linear cluster state is confirmed to be fully inseparable,
the full inseparability of the square cluster state is verified at the same time,
as the square and linear cluster states are locally equivalent
up to local Fourier transforms.
For the linear cluster state, we obtain
\begin{center}
\begin{equation}
\begin{split}
\langle[\Delta(\hat{p}_{\mathrm{L}1}-\hat{x}_{\mathrm{L}2})]^{2}\rangle+\langle[\Delta(\hat{p}_{\mathrm{L}2}-\hat{x}_{\mathrm{L}1}-\hat{x}_{\mathrm{L}3})]^{2}\rangle < 1\\
\langle[\Delta(\hat{p}_{\mathrm{L}3}-\hat{x}_{\mathrm{L}2}-\hat{x}_{\mathrm{L}4})]^{2}\rangle+\langle[\Delta(\hat{p}_{\mathrm{L}2}-\hat{x}_{\mathrm{L}1}-\hat{x}_{\mathrm{L}3})]^{2}\rangle < 1\\
\langle[\Delta(\hat{p}_{\mathrm{L}3}-\hat{x}_{\mathrm{L}2}-\hat{x}_{\mathrm{L}4})]^{2}\rangle+\langle[\Delta(\hat{p}_{\mathrm{L}4}-\hat{x}_{\mathrm{L}3})]^{2}\rangle < 1.
\end{split}\label{LinearInseparablity}
\end{equation}
\end{center}



For the T-shape cluster state, we have
\begin{center}
\begin{equation}
\begin{split}
\langle[\Delta(\hat{p}_{\mathrm{T}2}-\hat{x}_{\mathrm{T}1})]^{2}\rangle+\langle[\Delta(\hat{p}_{\mathrm{T}1}-\hat{x}_{\mathrm{T}2}-\hat{x}_{\mathrm{T}3}-\hat{x}_{\mathrm{T}4})]^{2}\rangle < 1\\
\langle[\Delta(\hat{p}_{\mathrm{T}3}-\hat{x}_{\mathrm{T}1})]^{2}\rangle+\langle[\Delta(\hat{p}_{\mathrm{T}1}-\hat{x}_{\mathrm{T}2}-\hat{x}_{\mathrm{T}3}-\hat{x}_{\mathrm{T}4})]^{2}\rangle < 1\\
\langle[\Delta(\hat{p}_{\mathrm{T}4}-\hat{x}_{\mathrm{T}1})]^{2}\rangle+\langle[\Delta(\hat{p}_{\mathrm{T}1}-\hat{x}_{\mathrm{T}2}-\hat{x}_{\mathrm{T}3}-\hat{x}_{\mathrm{T}4})]^{2}\rangle < 1.
\end{split}\label{T-shapeInseparablity}
\end{equation}
\end{center}

Note that the variances of a vacuum state are $\langle[\Delta\hat{x}_{\mathrm{vac}}]^{2}\rangle=\langle[\Delta\hat{p}_{\mathrm{vac}}]^{2}\rangle=\frac{1}{4}$.

The values of the left-hand-sides of the inequalities are
$0.34\pm0.02$, $0.42\pm0.02$, and $0.35\pm0.02$ for Eq.~(\ref{LinearInseparablity}), respectively,
and $0.42\pm0.02$, $0.43\pm0.02$, and $0.42\pm0.02$ for Eq.~(\ref{T-shapeInseparablity}), respectively.
Thus, all inequalities are simultaneously satisfied and hence
the full inseparability of the created cluster states is verified \cite{Inseparability}.

Besides confirming the inseparability of the cluster states,
we also verified that the measured correlations correspond to the squeezing levels
of the input states.
It is possible to detect the squeezing levels by removing the beam splitters.
The measured squeezing levels range from $-5.5$dB to $-6.3$dB and the antisqueezing levels
are between $+9.1$dB and $+11.9$dB.
After removing the beam splitters needed for generating the cluster states,
the signal at this stage of the experiment is free of fluctuations in phase locking.
Therefore, the squeezing levels are slightly better than the measured correlations.
Nonetheless, our results demonstrate the efficient generation
of the desired quadrature quantum correlations through
cancellation of all antisqueezing components of the light fields involved.

\section{Conclusion}

In conclusion, we have demonstrated
the generation of three kinds of CV four-mode cluster states.
Our method for cluster-state generation is very efficient,
because it completely suppresses the presence of antisqueezing components
in the output states. Thus, the input squeezing levels are almost
perfectly transferred onto the outgoing quadrature quantum correlations.

The approach here can be easily extended to other cluster states,
including higher numbers of modes and more complex graph states
suitable for multi-mode, multi-step quantum operations.
In order to perform a specific quantum operation on an input state through one of the cluster states,
the input state must be attached to that cluster state.
This could be done for arbitrary input states using a QND gate from off-line squeezing.
Such a QND gate has been experimentally realized already \cite{QNDExp}.
For processing the input state,
measurements have to be performed on most of the modes of the cluster state.
For universal quantum gates, non-Gaussian measurements are
needed and, in addition, feedforward operations
including active, conditional basis changes between the measurement steps.

With the linear cluster state, single-mode four-step operations are possible.
The same state can be also used as a horseshoe cluster for two-mode quantum operations \cite{Walther05}.
With the square cluster state, two-mode quantum operations can be performed,
and by employing it as a diamond state, it can be used in a redundant encoding scheme
for error filtration \cite{Peter07}.

\vspace{3mm}
This work was partly supported by SCF and GIA commissioned by the MEXT of Japan, and Research Foundation for Opto-Science and Technology.
PvL acknowledges support from the DFG under the Emmy Noether programme.
The authors acknowledge Seiji Armstrong for assistance with the editing of this report.

\end{document}